\documentclass[conference]{IEEEtran}
\IEEEoverridecommandlockouts
\usepackage{cite}
\usepackage{amsmath,amssymb,amsfonts,amsthm,bm}
\usepackage{algorithmic}
\usepackage{graphicx}
\usepackage{textcomp}
\usepackage{xcolor}
\usepackage{booktabs}
\usepackage{gensymb}
\usepackage{hyperref}
\usepackage{balance}
\def\BibTeX{{\rm B\kern-.05em{\sc i\kern-.025em b}\kern-.08em
    T\kern-.1667em\lower.7ex\hbox{E}\kern-.125emX}}

\begin{document}

\IEEEpubid{\begin{minipage}[t]{\textwidth}\ \\[10pt]
        \centering\normalsize{979-8-3503-1104-4/23/ \copyright 2023 IEEE}
\end{minipage}} 

\title{Reconstructing the Dynamic Directivity \\of Unconstrained Speech
\thanks{This research was funded by Meta Reality Labs Research through a summer internship and extended academic collaboration.}
}

\author{
\IEEEauthorblockN{Camille Noufi}
\IEEEauthorblockA{\textit{Center for Computer Research in Music and Acoustics} \\
\textit{Stanford University}\\
Stanford, CA, USA \\
cnoufi@ccrma.stanford.edu}

\and

\IEEEauthorblockN{%
Dejan Markovic, 
Peter Dodds\textsuperscript{\textsection}
}%
\IEEEauthorblockA{\textit{Reality Labs Research} \\
\textit{Meta Inc.}\\
Redmond, WA, USA \\
dejanmarkovic@meta.com, phdodds@gmail.com}
}


\maketitle
\begingroup\renewcommand\thefootnote{\textsection}
\footnotetext{Work conducted while employed at Meta Reality Labs Research.} 
\endgroup

\begin{abstract} 
This article presents a method for estimating and reconstructing the spatial energy distribution pattern of natural speech, which is crucial for achieving realistic vocal presence in virtual communication settings. The method comprises two stages. First, recordings of speech captured by a real, static microphone array are used to create an egocentric virtual array that tracks the movement of the speaker over time. This virtual array is used to measure and encode the high-resolution directivity pattern of the speech signal as it evolves dynamically with natural speech and movement. In the second stage, the encoded directivity representation is utilized to train a machine learning model that can estimate the full, dynamic directivity pattern given a limited set of speech signals, such as those recorded using the microphones on a head-mounted display. Our results show that neural networks can accurately estimate the full directivity pattern of natural, unconstrained speech from limited information. The proposed method for estimating and reconstructing the spatial energy distribution pattern of natural speech, along with the evaluation of various machine learning models and training paradigms, provides an important contribution to the development of realistic vocal presence in virtual communication settings. 
\end{abstract}


\section{Introduction}\label{sec:intro}

As virtual methods of communication continue to gain prevalence, achieving a sense of vocal presence within an augmented or virtual reality (AR/VR) paradigm that is natural and indistinguishable from reality has become increasingly necessary. Research has suggested that realistic propagation of the voice through a virtual environment often enhances this sense of vocal presence~\cite{Wendt2018,Ehret2020,Kato2010,Postma2016,Postma2017}. Therefore, it is crucial to be able both to understand and to emulate how a natural speech signal radiates from the mouth and reflects off the body into space.

Speech is a time-varying and frequency-dependent acoustic signal. It consists of a continuous range of frequencies from approximately 100 Hz (in low register voices) up to 16 k Hz. As a signal's wavelength decreases relative to the size of its source, its radiation pattern becomes increasingly directional~\cite{Beranek}. Thus, speech results in a dynamic frequency-dependent combination of radiation patterns ranging from omnidirectional to highly directional~\cite{Lindblom2007}. Moreover, in natural communication, tonal modifications, physicality, and body language constantly vary in reaction to context, resulting in additional changes to the directivity pattern. While speech directivity has been studied for a long time, expected directivity patterns have been primarily established from measuring segmented parts or modes of speech. In contrast, this study focuses on the directivity patterns of speech during natural movement and communication.

In this article, we introduce a method that utilizes deep learning to reconstruct the dynamic directivity of unconstrained speech. Specifically, in Section \ref{sec:theory}, we propose a virtual, egocentric microphone array that actively tracks the speaker's movements, enabling us to capture highly detailed directivity patterns of natural speech. Next, we put forth several variations of machine-learning models to accurately estimate the aforementioned patterns, given a limited set of speech signals. The detailed implementation of our directivity capture system and machine-learning experiments can be found in Section \ref{sec:experiments}. In Section \ref{sec:est_results}, we present our results, which demonstrate the successful regression from a limited set of speech signals to a high-resolution, dynamic directivity pattern. We discuss our method and propose future work in Section \ref{sec:Discussion}, and review our contributions in Section \ref{sec:conclusion}.

\subsection{Related Work}\label{sec:related}

Previous research on source and listener directivity in virtual communication provides support for the importance of accurate speech directivity rendering \cite{Mehra2014, Gotz2019}. In virtual and augmented communication, reproducing the talker's speech directivity pattern has been found to be crucial for an authentic conversational experience \cite{Wendt2018, Ehret2020}. Accurate speech directivity helps the listener determine the speaker's facing direction \cite{Kato2010} and increases the realism of dynamic speech \cite{Postma2016, Postma2017}, bodily situational awareness and movement perception from the listener's perspective \cite{Stanton2020}.

Historically, speech directivity studies have focused on capturing acoustic radiation patterns as a function of vocalization type. Research has repeatedly confirmed that the directivity of an acoustic signal becomes increasingly narrow as the signal's spectral contents increase in frequency \cite{Flanagan1960, Marshall1985, Kob1999, Warnock2002, Best2005, Monson2012}. Thus, directivity patterns of different vowels and consonants, consisting of varying spectral composition, produce different radiation patterns. Several studies have analyzed the directivity of held or isolated vowels and consonants extracted from speech \cite{Monson2012,Katz2006, Kocon2018, Blandin2018, Porschmann2020} and singing \cite{Katz2007}. Furthermore, studies have looked at the effects of vocal tract and mouth configuration \cite{Blandin2018, Bozzoli2005, Blandin2019, Brandner2020}, vocal technique \cite{Monson2012, Katz2007}, projection and loudness \cite{Marshall1985, Warnock2002, Monson2012, Katz2007}, gender \cite{Monson2012, Bellows2019}, interpersonal deviation \cite{Bellows2019}, hand position~\cite{porschmann2022effects} and torso configuration \cite{Brandner2020}. Although loudness and projection mode have not been found to significantly modify the directivity, the results of these studies suggest that vocal technique and singing style, which require specific modifications to mouth shape, alter the spectrum of the voice signal and, in turn, the directivity. Studies on instrument directivity have shown a firsthand effect of timbre and instrument body shape on the resulting directivity pattern \cite{Behler2016, Shabtai2017, Bodon2015, Canclini2019}. Most of the aforementioned studies have analyzed directivity patterns within frequency bands such as 1/3-octave or 1-octave bands. Recent studies have provided evidence arguing for the inclusion of frequency content up to 16 k Hz when studying or modeling voice directivity \cite{Monson2012, Kocon2018}.

Research in the field of voice directivity patterns has largely relied on simulation or physical measurement through the use of microphone arrays. For physical measurement, microphones are placed at predetermined locations on circular, spherical, or hemispherical arrays to capture the voice at specific points in physical space. The signals captured at these points are then combined to determine the directivity pattern of the signal. 
Over time, numerous configurations have been developed to capture either circular or spherical radiation patterns. Several studies have focused on azimuthal directivity using simplified setups, such as a horizontal ring array. This has been achieved by placing a subject inside a semicircular array positioned at mouth-height and capturing directivity in a single recording (single capture)~\cite{Monson2012, Kocon2018}, or by placing the subject on a turntable and rotating the turntable by a fixed amount while the subject repeats a vocalization (multiple capture)~\cite{Kob1999, Warnock2002}. Other studies have used a rotating semicircular vertical and/or horizontal array to record the sound pressure level (SPL) at locations offset in both azimuth and elevation~\cite{Warnock2002,Katz2006, Katz2007,  Bozzoli2005, Bellows2019}. The spatial resolution in these studies ranged from 5\degree\ to 20\degree\ in azimuth or elevation. A few studies have enclosed a stationary speaker inside of a spherical array, constructed of either azimuthal rings with microphones at a specified spatial resolution (often referred to as a \textit{regular grid})~\cite{Marshall1985}, or a more sparse but evenly-distributed spherical spacing~\cite{Bellows2019}. These captures are generally converted to directivity measurements by calculating the difference between the SPL at the location of interest and either a reference microphone near the mouth~\cite{Warnock2002, Katz2007, Bozzoli2005,Bellows2019} or the average omnidirectional SPL~\cite{Monson2012, Kocon2018}. 
A recent study aimed to overcome some of the limitations of the aforementioned studies by measuring human speech directivity with high resolution over a complete sphere using live phonetically-balanced passages. The study used a multiple-capture transfer function technique and spherical harmonic expansions to produce new directivity patterns~\cite{leishman2021high}.

In contrast to previous works focusing on capturing directivity of specific vocalizations, our proposed method introduces a novel approach to capture \textit{and} reconstruct the dynamic directivity of unconstrained speech. Rather than constraining our speaker to obtain physical directivity measurements, we concentrate on developing a virtual, egocentric microphone array method that accurately tracks with the speaker's movements. To the best of our knowledge, the estimation and reconstruction of these dynamic, high-resolution patterns has not been previously attempted. 
 
\section{Two-Stage Method for Directivity Measurement and Estimation}\label{sec:theory}

\begin{figure*}[htb!]
  \centering
  \includegraphics[width=0.95\textwidth]{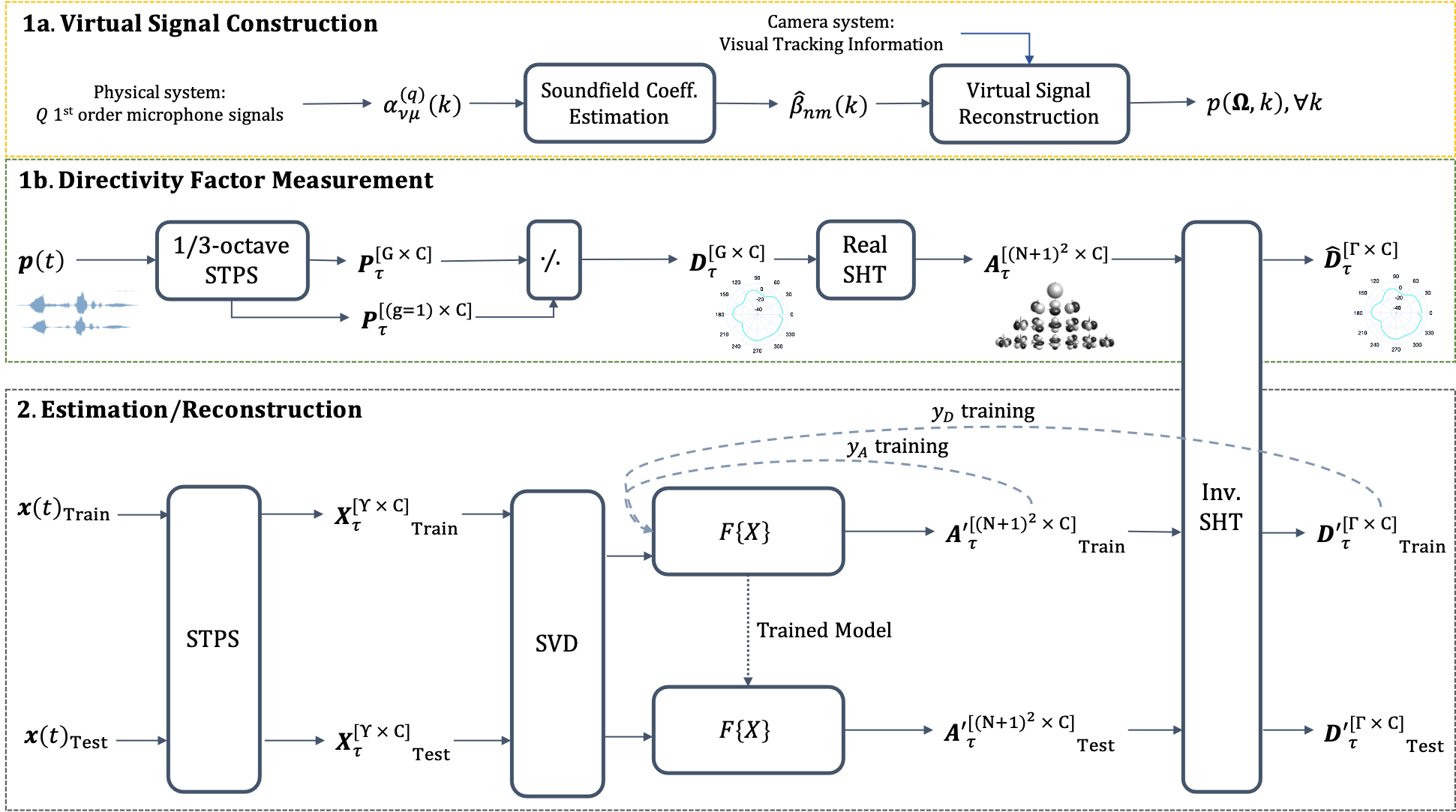}
  \caption{
  Reconstruction Framework: Proposed pipeline for capturing and estimating high-resolution speech directivity. Part 1a: Constructing Signal $p$ for $G$ points $\bm{\Omega}_{G}$ on an egocentric virtual microphone array, described by equations Eq. \eqref{eq:vmic_signal} to Eq. \eqref{eq:translation}. Part 1b: Measuring the empirical directivity factor $\bm{D}$ using the 1/3-octave short-time power spectrum (STPS) and calculating the analytical directivity factor $\bm{\Hat{D}}$ using the spherical harmonic transform (SHT), as described by equations Eq. \eqref{eq:df} to Eq. \eqref{eq:tm}. Part 2: Reconstructing the high-resolution directivity factor $\bm{D^\prime}$ from a limited set of signals $\bm{x}$ recorded near the mouth. This is achieved by predicting the spherical harmonic coefficients $\bm{A}$ using a regression model $F\{\bm{X}\}$, as explained in Section \ref{sec:est_theory}.
}
  \label{fig:endtoend}
\end{figure*}

In this section, we present our proposed framework for measuring and reconstructing the time-varying egocentric directivity pattern of natural unconstrained speech. We outline a method of creating the egocentric directivity pattern from a speaker who freely moves and speaks within a capture chamber. This pattern is encoded into short-time, frequency-dependent, real spherical harmonic coefficients using the spherical harmonic representation \cite{williams1999fourier} (Section \ref{sec:df_theory}), which serves as the basis for learning a data-driven regressive mapping from limited signal information to a high-resolution approximation of the measured pattern (Section \ref{sec:est_theory}). The two-stage capture and reconstruction framework is illustrated in Figure \ref{fig:endtoend}.

\subsection{Measurement of Egocentric Directivity Pattern}\label{sec:measurement_framework}

\subsubsection{Egocentric Virtual Microphone Array}\label{sec:real2virtual}

Here we consider a scenario in which the only source of sound, e.g., a human speaker, is found inside an enclosed capture area, contained within a smaller region of radius $R$ (see Figure \ref{fig:capture chamber}). In this case the virtual microphone derivation follows the approach outlined in \cite{SndFldRec3D_2012}\footnote{For more comprehensive analyses and solutions for scenarios with outside sources of sound and reverberation see \cite{SndFldSep_2017, SndFldRec_icassp2019, SndFldRec_waspaa2019}.}.

The derivation of virtual microphone signals at arbitrary points in space is formulated in the spherical harmonics domain.
Note that the derivation uses short-time analysis to account for dynamic scenarios but frame index is omitted for clarity of notation.
Let $k = 2\pi f/c$ be the wave number, $f$ be the frequency in Hertz, and $c$ indicate the speed of sound in meters per second. The acoustic signal $p(\bm{\Omega},k)$ at point $\bm{\Omega} = (r,\theta,\phi)$ 
outside the source region is given by 
\begin{equation}
\small
\begin{split}
p(\bm{\Omega},k) = & \sum_{n=0}^{N = \lceil keR/2 \rceil}\sum_{m=-n}^n\beta_{nm}(k)h_n(kr)Y_{n}^m(\theta,\phi) \ ,\\
\end{split}
\label{eq:vmic_signal}
\end{equation} 
where $Y_{n}^m(\theta,\phi)$ represents the complex spherical harmonic of order $n$ and degree $m$, 
$h_n(\cdot)$ is $n$-th order spherical Hankel function, $\beta_{nm}(k)$ denotes the soundfield coefficients,
$\lceil \cdot \rceil$ is the ceiling operator and $e$ is the Euler's number. 
As a consequence, if the global coefficients $\beta_{nm}(k)$ are known up to the order $N$, 
the signal at any virtual microphone position can be reconstructed using Eq. \eqref{eq:vmic_signal}. 

We estimate $\beta_{nm}(k)$ from the signals captured by $Q$ real $V$th-order microphones surrounding the capture area. 
Each $V$th-order microphone captures local soundfield coefficients $\alpha^{(q)}_{\nu\mu}(k)$, $q=1,\ldots,Q$, 
up to the order $V$, i.e., $\nu=0,\ldots,V$, $ -\nu \leq \mu\leq \nu$. These coefficients describe the soundfield locally at the $q$th microphone
position $\bm{x}_q = (r_q,\theta_q,\phi_q)$. 
The relationship between the local coefficients $\alpha^{(q)}_{\nu\mu}(k)$ and the 
global coefficients ${\beta}_{nm}(k)$ in Eq. \eqref{eq:vmic_signal} is given by the 
addition theorem \cite{martin2006multiple}
\begin{equation}
\small
\begin{split}
\alpha^{(q)}_{\nu\mu}(k) = & \sum_{n=0}^{N}\sum_{m=-n}^n\beta_{nm}(k)S_{n\nu}^{m\mu}(\bm{x}_q,k)\ ,
\end{split}
\label{eq:additionTheorem}
\end{equation}
where
\begin{equation}
\footnotesize
\begin{split}
S_{n\nu}^{m\mu}(\bm{x}_q,k) =\ & 4\pi i^{\nu-n}\sum_{l=0}^{n+\nu}i^l (-1)^{2m-\mu}h_l(kr_q)Y_l^{(\mu-m)*}(\theta_q,\phi_q)\\ &\sqrt{(2n+1)(2\nu +1)(2l+1)/4\pi}W_1W_2 \ ,
\end{split}
\label{eq:translation}
\end{equation}
encodes the global to local coefficient translation, $i$ is the imaginary unit, $^*$ denotes the complex conjugate, and $W_1$ and $W_2$ are Wigner 3-$j$ symbols defined as in \cite{martin2006multiple}. We form a linear system using Eq. \eqref{eq:additionTheorem} for all $Q(V+1)^2$ local coefficients $\alpha^{(q)}_{\nu\mu}(k)$, and 
the solution of the inverse problem gives the estimates of the $(N+1)^2$ global coefficients $\hat{{\beta}}_{nm}(k)$. Finally, given global soundfield coefficients, the signals of dynamically moving virtual microphones are computed using Eq. \eqref{eq:vmic_signal} on a frame-by-frame basis, where microphone positions at each frame are obtained from the visual tracking of the speaker. This process is summarized as part 1a. in Figure \ref{fig:endtoend}.

\subsubsection{Analytical Directivity Pattern}\label{sec:df_theory}

To capture a high-resolution sampling of a speaker's speech radiation pattern in three dimensions, we place $G$ virtual microphones on a virtual spherical grid with fixed far-field radius $R$ from the speaker. The locations of these $G$ microphones are in the set $\bm{\Omega}_G = \{(\theta, \phi)_1, (\theta, \phi)_g \dots (\theta, \phi)_G\}$, where $\theta$ and $\phi$ are azimuth and elevation angles, respectively. The acoustic signals captured by each virtual microphone are in the set $\bm{p} = \{p_{1}(t), p_{g}(t)\dots p_{G}(t)\}$. For each signal $p_{g}(t)$, we compute its power spectrum $\bm{P}_{g}$ over $\tau \in T$ frames and $C$ 1/3-octave center frequencies $\bm{f}_c$ \cite{AcousticalSocietyofAmerica2004}. In accordance with the speech directivity measurement literature \cite{Warnock2002, Katz2007, Bozzoli2005,Bellows2019}, we define the directivity factor $D_{g}(\tau, f_{c})$ to be the ratio between the power spectrum $P_g(\tau, f_{c})$ at $\Omega_{g}=(\theta, \phi)_g$ and the power spectrum $P_1(\tau, f_{c})$ of the reference signal at $\Omega_{1}=(0,0)$:

\begin{equation}\label{eq:df}
\small
\begin{aligned}[t]
  D_g(\tau,f_c) = \frac{P_g(\tau,f_c)}{P_1(\tau,f_c)}. 
\end{aligned}
\end{equation}

The directivity factor definition provides us with a way to understand the directivity in the context of the application at hand. For example, when we have access to a reference microphone signal positioned in front of the speaker (which is often the case when using a head-mounted display), the directivity factor helps us measure the distribution of speech energy around the speaker relative to the energy at the reference microphone's location. A directivity factor of 1 signifies that the sound pressure level (SPL) at a specific location on the sphere is equal to that at the reference location $\Omega_1 = (0,0)$. Consequently, if the directivity factor $D_g(\tau,f_c)$ consistently equals 1 across a cross-sectional plane of the grid, it indicates an omnidirectional pattern on that plane. On the other hand, a directivity factor greater than 1 signifies that the SPL at that particular location is higher than the SPL at the reference point.

Spherical harmonics provide a versatile way to encode the measured directivity pattern. Given $G$ points distributed around a spherical array, we can use spherical harmonics up to a maximum order of $N \leq \sqrt{G}-1$ to encode the measurements. The encoded representation can be decoded using the inverse spherical harmonic transform (SHT), allowing us to sample the sphere at up to $(N+1)^2$ uniform points.

For each frame $\tau \in T$ and each center frequency $f_c$, we calculate the \textit{real} spherical harmonic coefficients
\begin{equation}\label{eq:sh_coeff}
\small
\begin{aligned}[b]
  a_{n,m}(\tau,f_{c}) = \sum_{g=1}^{G} {D}_{g}(\tau,f_{c}) Y_n^m(\Omega_{g})\xi_{g}
\end{aligned}
\end{equation}
using the real spherical harmonic basis functions
\begin{equation}\label{eq:sh_bases}
\footnotesize
\begin{aligned}[b]
    Y_n^m(\theta,\phi) = 
    \begin{cases}
        \sqrt{2}\sqrt{\frac{2n+1}{4\pi}\frac{(n-|m|)!}{(n+|m|)!}L_n^{|m|}\cos{\phi}\sin{(|m|\theta})},& m<0 \\
        \sqrt{\frac{2n+1}{4\pi}L_n^{|m|}}\cos{\phi},& m=0 \\
        \sqrt{2}\sqrt{\frac{2n+1}{4\pi}\frac{(n-|m|)!}{(n+|m|)!}L_n^{|m|}\cos{\phi}\cos{(m\theta})},& m>0
    \end{cases}
\end{aligned}
\end{equation}
where $L_n^{|m|}$ are the associated Legendre polynomials. Voronoi grid sampling weights $\xi_g$ are used when coordinates $\bm{\Omega}_G$ are sampled using a rectangular or other non-uniform grid. The analytical directivity factor $\Hat{D}_{(\theta,\phi)}$ is found via the inverse spherical harmonic transform (SHT):
\begin{equation}\label{eq:df_ana}
\begin{aligned}[b]
  \Hat{D}_{(\theta,\phi)}(\tau,f_c) = 
  \sum_{n=0}^{N}\sum_{m=-n}^{n} a_{n,m}(\tau,f_{c}) Y_n^m(\theta,\phi).
\end{aligned}
\end{equation}

If we constrain the decoding process to a fixed radius $R$ and coordinate set $\bm{\Omega}_{\Gamma}$, where $\Gamma \leq (N+1)^2$, we can represent the inverse SHT as the matrix operation

\begin{equation}\label{eq:tm}
\small
\begin{aligned}[b]
  \bm{\Hat{D}}_\tau = 
  \bm{T}_\tau \bm{A}_\tau, 
\end{aligned}
\end{equation}
where the set of spherical harmonic bases $\bm{T} \in \mathbb{R}^{\Gamma \times  (N+1)^2}$ is defined as a predetermined differentiable transformation matrix that operates on real spherical harmonic coefficient matrix $\bm{A} \in \mathbb{R}^{(N+1)^2 \times C}$. This process is summarized as Part 1b. in Figure \ref{fig:endtoend}.

We note that it is technically possible to combine parts 1a. and 1b. However, by keeping these stages separate, we can explore the impact of different virtual array geometries on directivity estimation as well as remain agnostic to the physical system and equipment used. Additionally, the availability of egocentric virtual microphone signals produced in Part 1a. opens up possibilities for alternative directivity representations such as direct signal prediction.

\subsection{Estimation from a Limited Set of Signals}\label{sec:est_theory}

In the second stage of our approach, we focus on mapping a small number of acoustic signals captured near the speaker's mouth to the analytical spherical directivity pattern. This step is depicted as Part 2 in Figure \ref{fig:endtoend}.

\subsubsection{Representation of Limited Set of Signals}

We employ the method described in Section \ref{sec:real2virtual} to capture $\Upsilon$ speech signals, denoted as $\{x_{1}(t), x_{\upsilon}(t),\dots, x_{\Upsilon}(t)\}$, at fixed radius $\rho<R$ with corresponding coordinates $\bm{\Omega}_{\Upsilon} = \{(\theta, \phi)_1, (\theta, \phi)_\upsilon \dots (\theta, \phi)_{\Upsilon}\}$. It is assumed, though not strictly required, that $\Upsilon$ is significantly smaller than $G$, and the distribution on the sphere is non-uniform. For each signal $x_{\upsilon}(t)$, we compute its power spectral density (PSD) $\bm{X}{\upsilon}(\tau)$ and create an aggregate PSD representation $\bm{X}(\tau) = \lbrack\bm{X}_1(\tau), \bm{X}_{\upsilon}(\tau), \dots, \bm{X}_{\Upsilon}(\tau)\rbrack$ encompassing all $\Upsilon$ virtual signals. We then apply truncated singular-value decomposition (SVD) to obtain a low-rank approximation of $\bm{X}(\tau)$. Further details regarding the implementation of SVD can be found in Section \ref{sec:experiments}. Subsequently, we utilize this compressed spatial PSD representation $\bm{X}$ as the input to our regression model.

\subsubsection{Model Variants}
\label{sec:architectures}
We compare a feed-forward multilayer perceptron network (MLP) \cite{Goodfellow-et-al-2016} and an encoder-decoder architecture with a recurrent encoder \cite{Hochreiter1997}, to several baselines. The baselines we select are three linear regression models (Ordinary least squares, Lasso, and Ridge) as well as naive-mean and naive-median predictions as baselines. To establish the notation for later use, we present the definition of a simple linear regression model and its training process:
\begin{equation}\label{eq:general_regression_func}
\begin{aligned}[b]
{\bm{y}^\prime}= F{\{\bm{X}}\} = \bm{w}^{T}\bm{X} + \bm{b},
\end{aligned}
\end{equation}
where $\bm{w}$ represents the weighting coefficients applied to each dimension of $\bm{X}$, and $\bm{b}$ is the bias term. The goal of the regression model $F{\{\bm{X}}\}$ is to learn the optimal weights $\bm{w}$ that provide the most accurate estimation $\bm{y^\prime}$ of the target directivity pattern $\bm{y}$ by minimizing the Euclidean distance (also known as the $\ell2$-norm):
\begin{equation}\label{eq:sh_cost_func}
\begin{aligned}[b]
d(\bm{y}, \bm{y^\prime}) = ||\bm{y} - \bm{y^\prime}||^{2}_{2}.
\end{aligned}
\end{equation}
This cost function quantifies the discrepancy between the target and estimated directivity patterns.

The MLP contains 7 fully connected hidden layers. The number of nodes per layer increases progressively, and each hidden layer is followed by a leaky rectified linear unit with a negative slope of 1e-2 \cite{Maas2013}. The final layer outputs a continuous vector $\bm{y}^\prime$ that estimates the flattened target sample $\bm{A}_\tau$ in one dimension. 


We design the encoder-decoder using an LSTM encoder and MLP decoder. This model predicts a sample $\bm{A}\tau$ given an input sequence $[\bm{X}_{\tau-s}, \dots, \bm{X}_{\tau-1}, \bm{X}_{\tau}]$, where $s$ is the sequence length in frames. The LSTM encoder consists of 5 layers with a hidden layer size of 64. The output of the last hidden layer at timestep $\tau$ yields a latent space that captures information from both the current frame $\tau$ and the previous $s$ frames. This latent space is then fed into the fully connected decoder, which is a 6-layer MLP with an output size of $C(N+1)^2$.


\subsubsection{Domain-Inspired Objective Functions}
\begin{table}[tb!]
\centering
\caption{Variations to objective function $d(\bm{y}, \bm{y^\prime})$. Four different specifications of input $\bm{y}$ and one variation to $d(\bm{y}, \bm{y^\prime})$ are considered. Frequency-weight matrix $\bm{W}$ and sign-weight function $s(\cdot)$ are defined in Eq. \eqref{eq:weighting_matrix} and Eq. \eqref{eq:sw_cost_func}, respectively.}
\label{tab:objective-funcs}
\begin{tabular}{@{}ll@{}}
\toprule
 & Input to Objective Function $d(\bm{y}, \bm{y^\prime})$ \\ \midrule
$\bm{y}_{A}$ & $\bm{A} = F(\bm{X})$ \\
$\bm{y}_{D}$ & $\bm{\hat{D}} = \bm{T}F(\bm{X})$ \\
$\bm{y}_{e}$ & $\bm{y} - \bm{\bar{y}}$ \\
$\bm{y}_{fw}$ & $(\bm{W}\bm{y}^{T})^{T}$ \\ \midrule
 & Objective Function Modification \\ \midrule
$d_{sw}$ & $\frac{||\bm{y}-\bm{y^\prime}||^{2}_{2}} {s(\bm{y},\bm{y^\prime})}$ \\ \bottomrule
\end{tabular}
\end{table}
We experiment with modifying the objective function in Eq. \eqref{eq:sh_cost_func} and explore several options, which are summarized in Table \ref{tab:objective-funcs}. Firstly, we investigate variations of the representation domain $\bm{y}$ that the objective function is applied to. Since the transformation matrix $\bm{T}$ is predetermined and differentiable, we can choose to decode to $\bm{\Hat{D}}$ (see Eq. \eqref{eq:tm}) either before or after minimizing Eq. \eqref{eq:sh_cost_func}. This choice changes the representation space that we optimize over. Decoding to $\bm{D}$ separately from computing the loss removes the decoding step from the training loop, resulting in $\bm{y}A$ in Table \ref{tab:objective-funcs}. This training scheme is useful if we require direct access to the estimated spherical harmonic coefficients $\bm{A}^\prime$ for downstream processing. Alternatively, we can include Eq. \eqref{eq:tm} in the forward and backward pass of the training process if we want to decode to a predetermined coordinate set $\bm{\Omega}\Gamma$, resulting in $\bm{y}_D$ in Table \ref{tab:objective-funcs}.

Furthermore, we consider predicting the deviation of a specific sample's directivity pattern from the expected value of the directivity pattern $\bm{E[y]}$. In this variant, we calculate the expected value of the target directivity pattern $\bm{y}$ across all frames as $\bm{E[y]}$, and train a model to predict the difference from the mean $\bm{y}_e = \bm{y} - \bm{E[y]}$. We apply this variation to discourage the model from predicting the mean and to focus on learning the variation between patterns.

Additionally, we explore the potential benefit of frequency band weighting for the objective function input ($\bm{y}_{fw}$ in Table \ref{tab:objective-funcs}). Since speech signals typically have most of their energy between 100 Hz and 8 k Hz, we construct a $C \times C$ diagonal weighting matrix $\bm{W}$. The matrix applies a linear up-weighting to values at indices of $\bm{y}$ that correspond to lower frequency bands. 
We define the elements along the diagonal of $\bm{W}$ to be
\begin{equation}\label{eq:weighting_matrix}
\begin{aligned}[b]
  w_k:=\left[1+k\left(\frac{M-1}{C-1}\right)\right],k=0,\ldots,C-1,
\end{aligned}
\end{equation}
where $M>1$ is the maximum weight and $k$ is the frequency-bin index. This weighting increases the distance between lower-frequency elements, making the model focus on minimizing the distance between the low-frequency regions of the representation space. This, in turn, helps reduce the overall loss. 

Finally, we impose an additional constraint on the objective function by dividing the output of equation Eq. \eqref{eq:sh_cost_func} by a differentiable function that yields a scalar inversely proportional to the number of incorrect sign predictions:
\begin{equation}\label{eq:sw_cost_func}
\begin{aligned}[b]
  s(\bm{y},\bm{y^\prime}) = 1 + \epsilon - \frac{\sum_{i=1}^{C(N+1)^2} 1 + \text{sgn}(y_i)\text{sgn}(y_i^\prime)}{C(N+1)^2}.
\end{aligned}
\end{equation}
This constraint aims to address the issue of weight updates reacting to estimation errors based solely on magnitude, as the $\ell2$-norm is the only source of information. By using this constraint, the model is encouraged to output the correct phase of the spherical harmonics when the input to the objective function is $\bm{y}_A$ or to generate spherical harmonic coefficients that result in predictions $\bm{\hat{D}} > 0$ when the input to the objective function is $\bm{y}_D$. This modification is also applicable to input $\bm{y}_{fw}$ and is denoted as $d_{sw}$ in Table \ref{tab:objective-funcs}.

\begin{figure*}[htb!]
  \centering
  \includegraphics[width=0.95\textwidth]{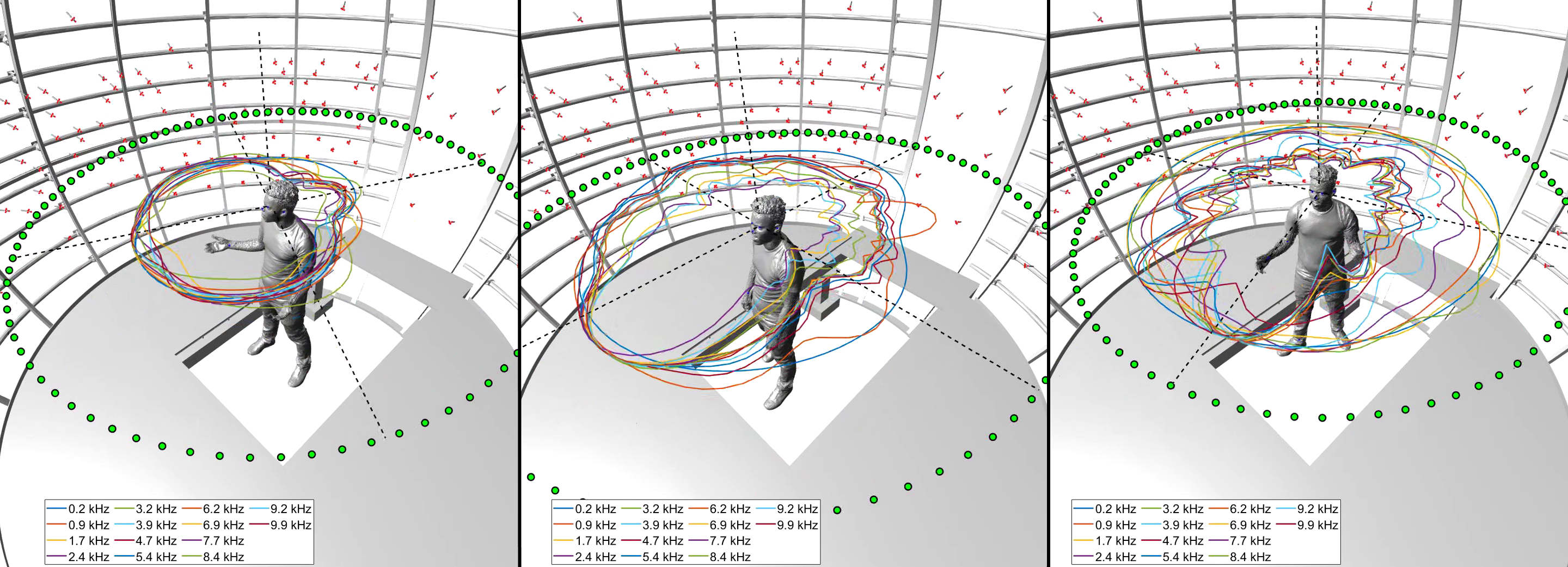}
  \caption{Example of egocentric virtual microphone array within the capture system. Red points are real microphones mounted on capture chamber walls. Green points are virtual locations of microphones spaced in a horizontal ring fixed to a 3-D coordinate system tracking with the speaker's position. As the speaker's mouth moves, turns and tilts in space, the coordinate system moves accordingly.}
  \label{fig:capture chamber}
\end{figure*}

\begin{figure}[h!]
\centering
\includegraphics[width=0.44\textwidth]{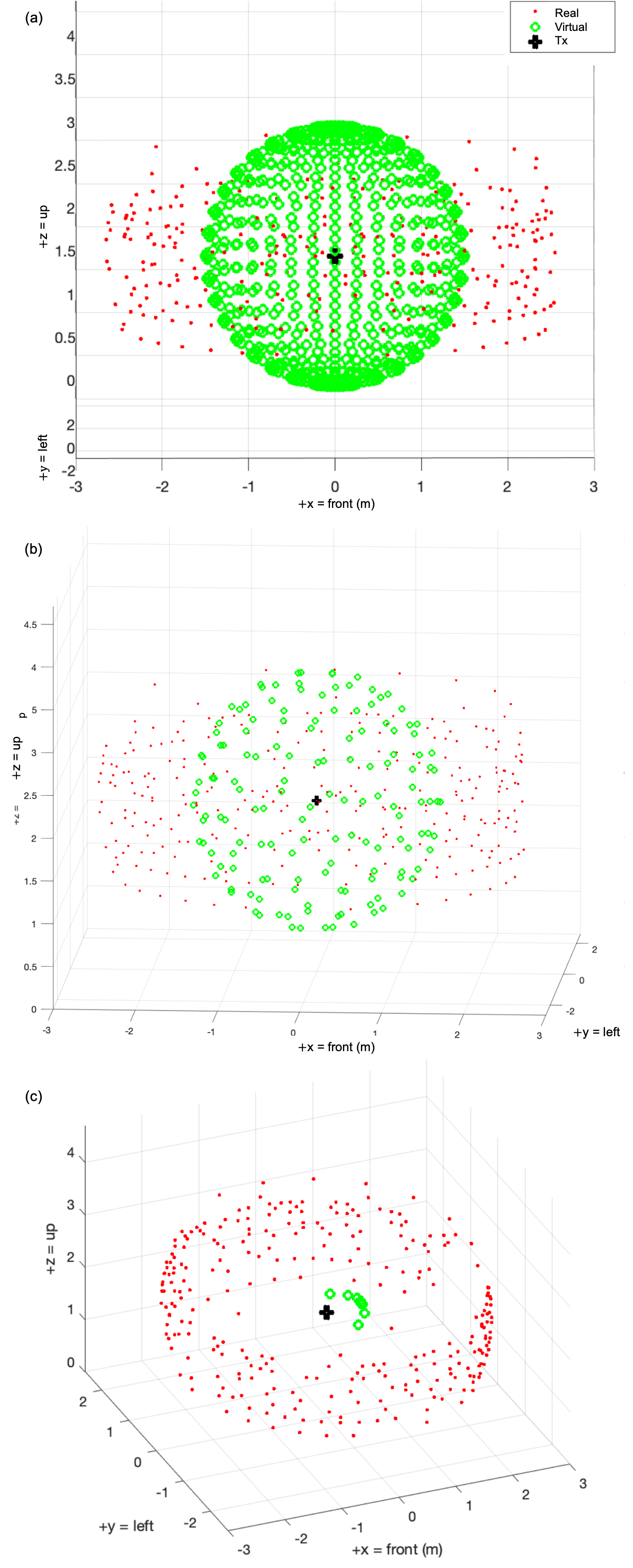}
\caption{Virtual spherical microphone array designs. \textit{$T_x$} denotes the speaker's mouth as a point source at the origin ($x=0$m, $y=0$m, $z=2$m). (a) 614-point Regular grid. (b) 144-point (10th order) \textit{t}-design grid. (c) Input array providing the proxy signals of the head-mounted display that are used in the estimation system. This proxy array consists of 8 microphones arranged in front of \textit{$T_x$} at a distance of 0.6m.}
\label{fig:grid-designs}
\end{figure}

\section{Experiments}\label{sec:experiments}

\subsection{Data Collection}
We implemented the system proposed in Section \ref{sec:df_theory} using a dataset comprising of two minutes of continuous conversational speech from a single male speaker ($T_x$). The speech was recorded in a closed spherical audiovisual capture chamber (shown in Figure \ref{fig:capture chamber}) that measures 2.74 meters in radius. The capture chamber is truncated by the ceiling and floor, and it contains 281 first-order ambisonic microphones mounted on its walls. The microphones cover 360 degrees in azimuth and 30 degrees in elevation above and below the speaker's mouth. The speech signals were recorded using real microphones with a sampling rate of 48 k Hz and bit depth of 24. The capture chamber also has a camera system to track the movements of the speaker. The capture chamber is located in a 7.6m x 7.5m x 5.2m acoustically treated room. Reverberation effects are minimal, thus are not considered within this work.

A small, egocentric virtual array with coordinates
\begin{align*}\bm{\Omega}_{\Upsilon} = \{(60\degree,\ 0\degree),\ (300\degree,0\degree),\ (30\degree,\ 5\degree),\ (330\degree,\ 5\degree), \\
(10\degree,\ 10\degree),(0\degree,\ 10\degree),(0\degree,\ 5\degree),(350\degree,\ 10\degree)\}
\end{align*}
and radius $\rho = 0.6$m, was created to serve as a proxy to the angular locations of microphones that would be found on an HMD (near-field effects were not considered). Figure \ref{fig:grid-designs}c displays this input grid. The signals rendered at these coordinates provided the input information to the estimation system.

\subsection{Virtual Array Design}

\subsubsection{Spherical Array Design}

We created two virtual spherical arrays: a regular grid and a 10th order \textit{t}-design grid \cite{Delsarte1977}. The regular grid consists of sets of horizontal rings stacked vertically, with spatial resolution defined by far-field radius $R$, azimuth angle $\Theta$, and elevation angle $\phi$. We selected $R$ to be 1.5 meters and $\Theta$ and $\phi$ to be 10\degree\ for a balance between high-resolution capture, computation time, and comparison with previous literature. A single microphone was placed at each point, resulting in 614 virtual microphones in the regular grid. The \textit{t}-design grid was also created with a radius of 1.5 meters and contained 144 virtual microphones. The $\Theta$ and $\phi$ coordinates were determined using the lookup table in the Spherical Harmonics Toolbox for MATLAB \cite{SHToolbox}. These grids are illustrated in Figure \ref{fig:grid-designs}a and \ref{fig:grid-designs}b, respectively.

We used the method described in Section \ref{sec:real2virtual} to ensure that these egocentric spherical arrays maintained an origin consistent with the position and tilt of the speaker's mouth. Figure \ref{fig:capture chamber} shows an example of an egocentric virtual microphone array tracking with the speaker. The 144 virtual microphones in the \textit{t}-design grid were used to measure the directivity factor $\bm{D}$, which was then encoded into spherical harmonics using the method from Section \ref{sec:df_theory}. We used the regular grid coordinates as the target coordinates $\bm{\Omega}_\Gamma$ for model training and inference.

\subsubsection{Virtual Signal Validation}
\label{sec:preprocess}

Figure \ref{fig:grid-designs} shows that many of the virtual signals captured near the poles of the virtual array are above or below the highest and lowest real microphones mounted to the capture chamber walls. To ensure that we analyzed valid signals, we dynamically muted signals outside of the valid reconstruction space.\footnote{We emphasize that this step is only necessary due to the physical setup of our capture chamber. This step can be omitted if the physical microphones provide full spherical coverage.}
This valid space is bounded by the positions of the capture chamber's physical microphones. We located the angular positions of the 281 real microphones from the egocentric perspective of the spherical array's coordinate system (i.e., coordinate set $\bm{\Omega}_G$ was held constant) at 23 ms frames. For each frame, if a virtual microphone location was above or below the valid elevation range, audio signal at that microphone was attenuated by 60 dB. This dynamic muting method effectively set the signal's directivity factor to 0 for virtual coordinates outside of the valid reconstruction zone. Regular grid signals were thus more likely to be muted due to a larger portion of microphone locations being near the array poles. Approximately 2/3 of the regular grid signals were muted for a majority of the recording. In contrast, more than half of the \textit{t}-design grid signals were within the valid reconstruction zone region.

\subsection{Directivity Measurement}\label{sec:measure}

We computed the dynamic speech directivity factor from the signal set $\bm{p}$ using the method detailed in Section \ref{sec:df_theory}. Each signal $p_{g}(t)$ was processed to obtain the 1/3-octave-band power spectrum $\bm{P}_g$ over $C = 23$ frequency bins\footnote{We excluded frequency bands centered at 31.5 Hz and 63 Hz from analysis due to the absence of speech content in this range.}. This was achieved by applying a 2048-point FFT, a window size of 2048, and a hop size of 512. The directivity factor was determined at each frame using Eq. \eqref{eq:df}. Spherical-harmonic encoding and decoding were performed using the method described in \cite{SHToolbox} as outlined by eqs. Eq. \eqref{eq:sh_coeff} - Eq. \eqref{eq:df_ana}.

\begin{figure}[tb!]
  \centering
  \includegraphics[width=\columnwidth]{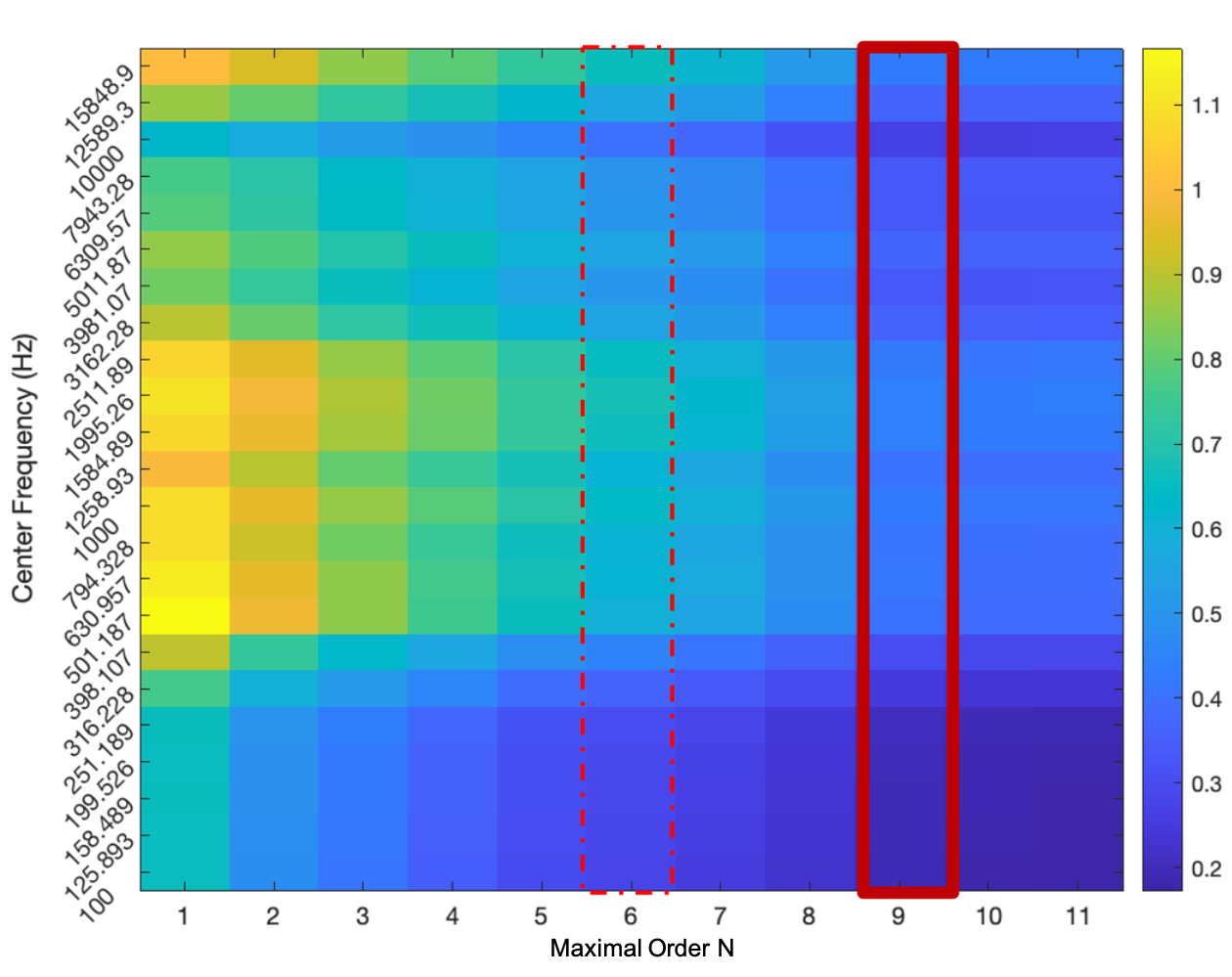}
  \caption{Reconstruction error (dB) between the measured and analytical directivity factor categorized by spherical harmonic order and frequency band. $N=9$ was selected to reduce error. $N=6$ is highlighted as an option for reducing the maximal order.}
  \label{fig:sh_order_df_error}
\end{figure}

We conducted a grid search over maximum spherical-harmonic orders $N \in [1,12]$ to identify the optimal maximum order that minimizes the reconstruction error between the measured and analytical directivity factors. We found that $N = 9$ was the optimal order to minimize reconstruction error (see Figure \ref{fig:sh_order_df_error}).

\subsection{Model Experiments}\label{sec:model_experiments}

\subsubsection{Preprocessing}

The training dataset consisted of only voiced frames, determined using the VOICEBOX energy-based active speech level detection algorithm \cite{Brookes2011}, resulting in 7,322 unique observations. Data augmentation was performed by applying additive white Gaussian noise at 10 dB and 20 dB signal-to-noise ratios to all input and target signals $\bm{p}_G$ and $\bm{x}$, respectively, before computing each signal's power spectral density. The speech recordings were partitioned into two separate audio regions. The first fifty seconds of each one-minute recording were used for training, while the final ten seconds were reserved for validation. The target dataset of spherical harmonic coefficients $\bm{A}$ was constructed from the signal set $\bm{p}_G$ measured at the \textit{t}-design points, while the decoding matrix $\bm{T}$ was predetermined for the 10\degree\ resolution regular grid. The power spectrum set $\bm{X}$ was constructed from the limited signal set $\bm{x}$ and compressed into its low-rank approximation using the ARPACK-SVD method \cite{ARPACK}. A rank of 50 was selected to achieve an explained variance ratio of 99\% across the training partition, and the same low-rank compression was applied to samples in the validation and inference partitions.

\subsubsection{Training}
The hyperparameters for each model were empirically tuned (hyperparameter optimization was not a focus of this study). For the Lasso regression model, we selected a regularization parameter of $\alpha=0.1$, while for the Ridge regression model, we chose $\alpha=0.9$. In all neural network models, batch normalization was applied after each linear layer, and a dropout rate of 20\% was used during training. The input sequence length $s$ to the LSTM encoder was set to 20 frames. A batch size of 64 was utilized, and the adaptive learning rate was initialized to 1e-3, decreasing by a factor of 10 after reaching the validation-set patience threshold of 5 epochs. Training was completed when the learning rate reached 1e-7. All neural network models were implemented in PyTorch and trained on a Tesla V100 GPU.

We used $\ell1$-loss to measure the directivity factor reconstruction error $\Delta \bm{D}$ because it is physically interpretable, although we trained the neural network using Euclidean distance as the objective function. To obtain $\bm{D}^\prime$, we added the expected value $\bm{E[y]}$ of the training partition to the predictions when $\bm{y}_e$ was fed into the objective function. When applying the frequency-weighted loss input $\bm{y}_{fw}$ and loss modification $d_{sw}(\cdot)$, we applied a weighting matrix $\bm{W}$ to the predictions and targets before feeding them into $d_{sw}(\cdot)$.

\section{Results}\label{sec:est_results}

\subsection{Directivity Measurement and Analytical Reconstruction}
\begin{figure}[htb!]
\centering
\includegraphics[width=0.95\columnwidth]{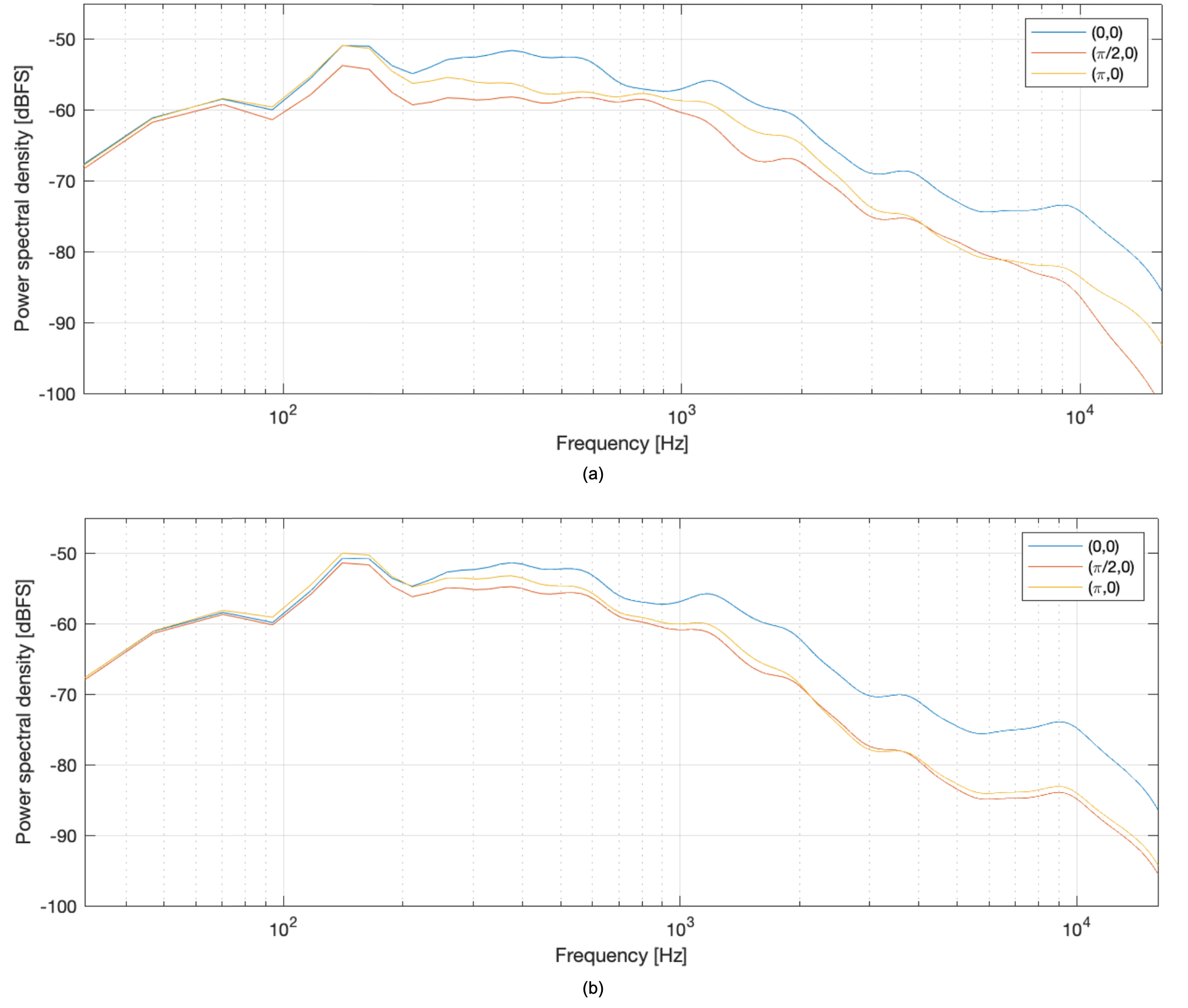}
\caption{Long-term average spectrum (LTAS) of signals captured from three directions: directly in front (0,0), to the left of ($\pi/2$,0), and behind ($\pi$,0) the speaker. (a) LTAS measured from the original signals. (b) LTAS obtained from the reconstructed signals using the short-time frame-wise estimates generated by the MLP with the loss input $\bm{y}_A$.}
\label{fig:ltas}
\end{figure}

\begin{figure*}[htb!]
\centering
  \includegraphics[width=0.85\textwidth]{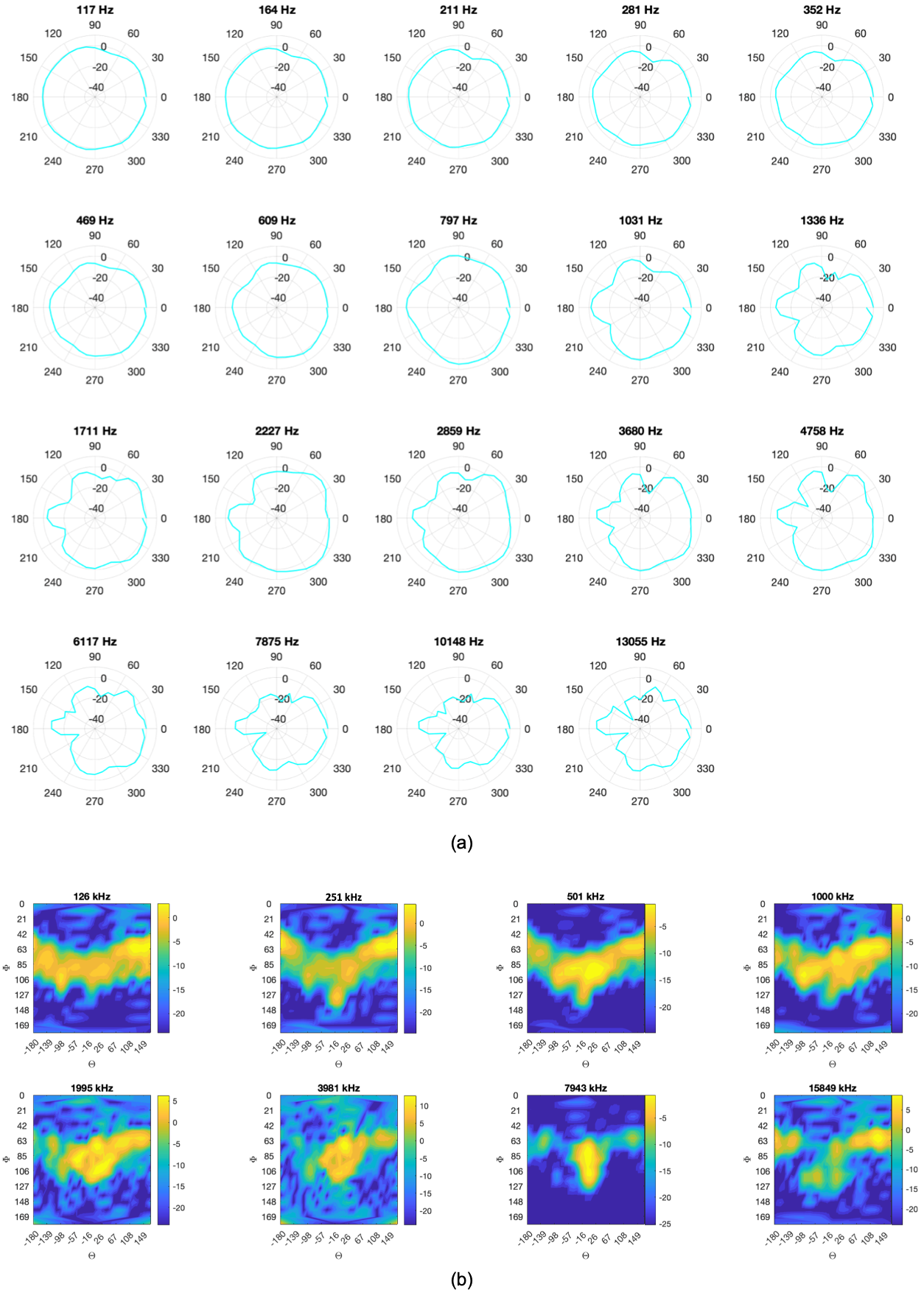}
  \caption{Frequency-dependent egocentric directivity index DI ($\log_{10}\bm{D}$) of a speech signal. (a) illustrates the relationship between directionality and frequency content throughout the signal duration. Each polar plot represents the DI of the running (LTAS) speech signal. The view is from the azimuthal plane at $\phi = \pi/2$, with the speaker's facing direction (0\degree) to the right. (b) provides an example of the DI for a single vowel extracted from the running speech. Each contour plot displays the average DI of the vowel $/aa/$ (5 manually-extracted instances) across the entire spherical array in each frequency band.}
\label{fig:analytical_df}
\end{figure*}

We calculate the average directivity of a running speech signal by applying Eq. \eqref{eq:df} to the long-term average spectrum (LTAS) of each signal $p_g(t)$. 
Figure \ref{fig:ltas}a displays the LTAS of signals directly in front of, to the left of, and behind the speaker. Figure \ref{fig:analytical_df}a presents polar plots in the azimuthal plane of the LTAS directivity index (DI).
As expected, these plots show that an increase in frequency leads to stronger directionality in front of the speaker's mouth. At very low frequencies ($<200$ Hz), the DI was omnidirectional. At low to mid-range frequencies, the DI attenuated by 5-10 dB. However, a prominent rear lobe emerged around 700 Hz alongside a slightly concave region near 70\degree. These features remained present up through high frequencies, as smaller side lobes and null regions became more prevalent between 60\degree\ and 300\degree. From 1 k Hz to 3 k Hz, the rear lobe edges occurred around -20 dB and attenuated to -40 dB at frequencies above 3 k Hz. These patterns were similar to those reported by Monson \textit{et. al} for frequencies up to 4 k Hz~\cite{Monson2012}, while we report a more pronounced rear lobe between 4 k Hz and 16 k Hz.

We look at the directivity pattern of the \textit{/aa/} vowel by manually extracting five occurrences from the running speech where the signal is pseudo-stationary and at least 80 ms in duration. We average the directivity factors of these five segments and plot the DI over the whole spherical array in Figure \ref{fig:analytical_df}b. Due to the necessary high and low elevation muting as described in Section \ref{sec:preprocess}, we are only able to comment on the pattern between 45\degree\ and 135\degree\ in elevation. We find that the directivity factor becomes increasingly narrow up to approximately 8 k Hz. Production of the \textit{/aa/} vowel produces little signal above 8 k Hz \cite{Monson2011}, so we do not expect to see meaningful directivity patterns at the 16 k Hz band.

\begin{figure*}[htb!]
\centering
 \includegraphics[width=0.85\textwidth]{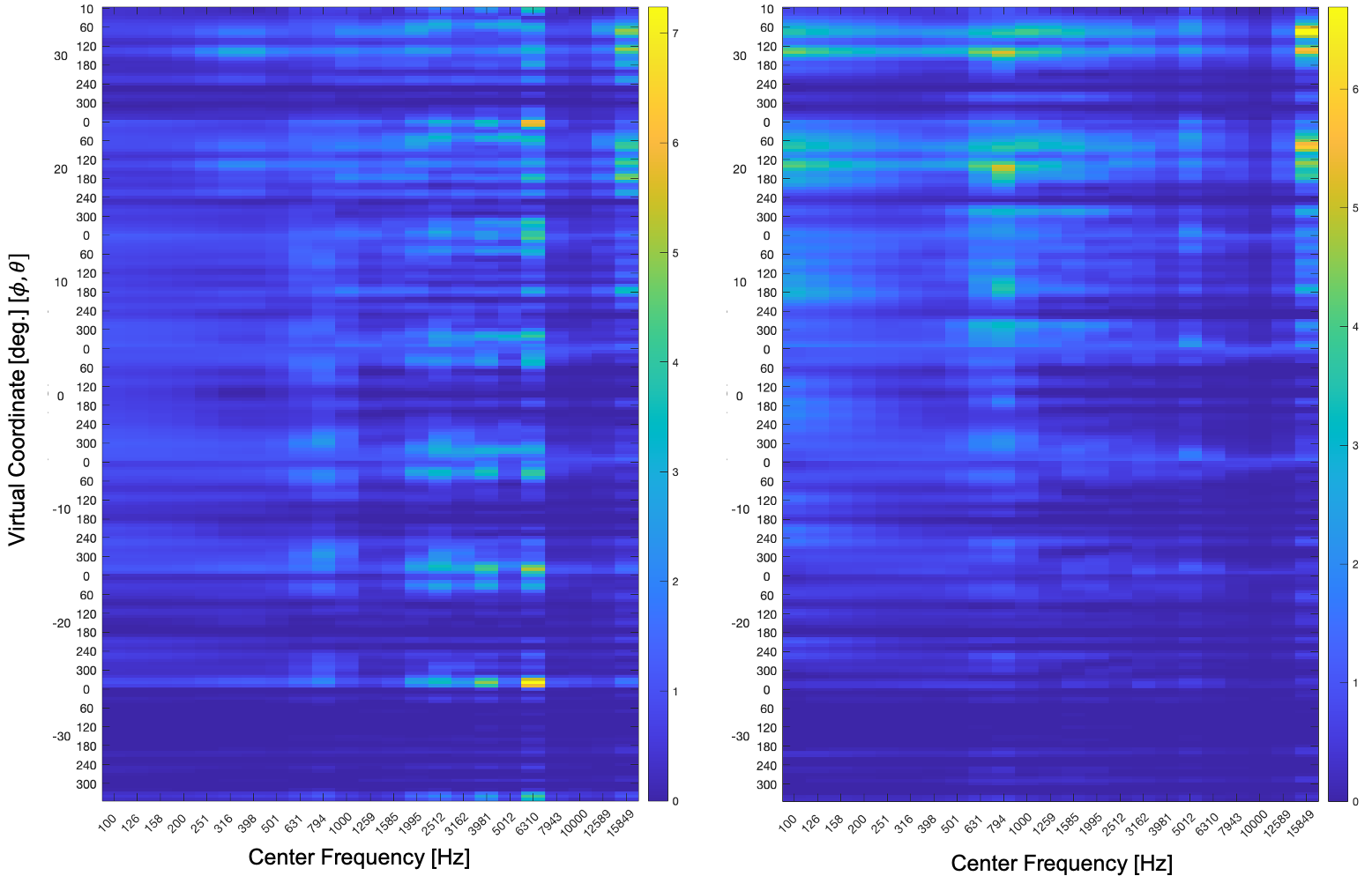}
\caption{Single-frame representations of the directivity factors for two different speech samples (SS1/left and SS2/right) from the same speaker. These directivity factors are reconstructed for all frequencies and virtual points $\Omega_{G}$ on a regular grid with a spacing of 10\degree.}
\label{fig:frame_error}
\end{figure*}

Additionally, we look at the analytical directivity factor $\hat{\bm{D}}_\tau$ of individual frames of voiced content. 
Figure \ref{fig:frame_error}a gives two examples (left/SS1 and right/SS2) of these single-frame directivity factors. Each plot shows $\hat{\bm{D}}_\tau$ at all points on the regular grid $\Omega_{G}$ from -30\degree\ to 30\degree\ in elevation. 
We note a few similarities between the two frames. At elevations below -20\degree\ we see a decrease in the directivity factor at most locations. An increase in directionality at higher-elevation areas of the grid occurs at lower frequency bands, suggesting that omnidirectionality does not necessarily always occur over whole sphere. Additionally, the reference location may not always be the focal point of the directional pattern. Consistent with the LTAS directivity index shown in Figure \ref{fig:analytical_df}a, we see an increase in directionality begin around 700 Hz. 
Although these general patterns occur across frames, we observe variations in magnitude at each grid location. One notable difference between these two frames is that SS2 exhibits a higher amount of energy distributed above the head, particularly in the low- and mid-range frequency bands, compared to SS1.

\subsection{Directivity Estimation}

\begin{table}[bt!]
\centering
\caption{Directivity Factor Reconstruction Error $\Delta \bm{D} = |\Hat{\bm{D}} - \bm{D}^{\prime}|$ over training and validation sets for models and baselines.}
\label{tab:results-model}
\begin{tabular}{@{}lccc@{}}
\toprule
 & Training & Validation & Inference Time\\ \midrule
\multicolumn{1}{l|}{LSTM} & 0.300 & 0.264 & 2.9 ms\\
\multicolumn{1}{l|}{MLP} & \textbf{0.222} & \textbf{0.200} & \textbf{2.1 ms}\\
\midrule
\multicolumn{1}{l|}{OLS} & 0.252 & 0.332 & -\\
\multicolumn{1}{l|}{Lasso} & 0.301 & 0.264 & -\\
\multicolumn{1}{l|}{Ridge} & 0.267 & 0.248 & -\\ \midrule
\multicolumn{1}{l|}{Mean} & 0.302 & 0.265 & -\\
\multicolumn{1}{l|}{Median} & 0.321 & 0.282 & -\\ \bottomrule
\end{tabular}
\end{table}

We used the reconstruction errors $\Delta \bm{D}$ to evaluate and compare the performance of each model in estimating high-resolution frame-wise directivity patterns using the limited signal set. Table \ref{tab:results-model} presents these errors. The naive-mean and naive-median prediction errors of the target directivity pattern $\bm{y}_A$ on the validation partition were 0.265 and 0.282, respectively. The validation errors were lower than the training errors, which can likely be attributed to the validation set having a smaller distribution of phonetic material and directivity patterns than the training set. The naive prediction errors suggested that any reconstruction error around or greater than 0.265 indicated a potentially non-functional mapping learned by the model. This outcome was observed for the LSTM, lasso, and ordinary least-squares regression models. The ridge regression model performed marginally better. The MLP network produced estimates with the lowest reconstruction error: 0.200 on the validation set. It predicted a target pattern quickly, with an inference time of 2.1 ms. The LSTM was marginally slower.

Frame-wise reconstruction errors ranged from 0-6 dB near and below the mouth. For lower frequencies, higher errors up to 12 dB were observed at elevations above the mouth. Above 3 k Hz, spurious errors up to 12 dB were observed across all elevations.

\begin{figure*}[htbp!]
  \centering
    \includegraphics[width=0.95\textwidth]{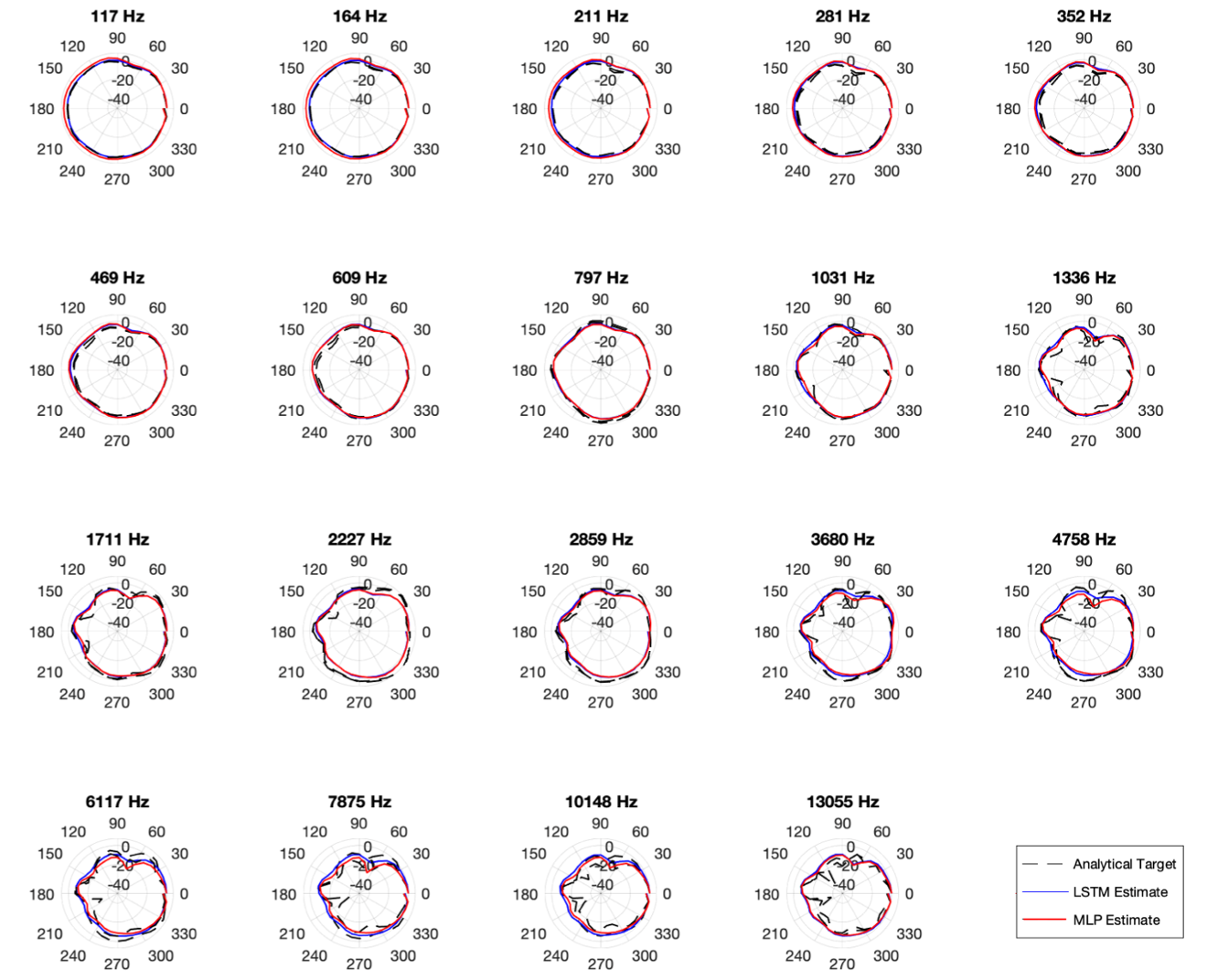}
  \caption{Directivity factor $\bm{D^\prime}$ of reconstructed audio estimated by the MLP (red) and LSTM (blue) models using loss input $\bm{y}_A$, as compared to the analytical directivity factor $\Hat{\bm{D}}$ (black, dashed). Shown here as directivity indices (polar plot scale in dB).}
  \label{fig:reconstruction-polar}
\end{figure*}

To assess the performance of our model in a more meaningful and perceptual manner, we reconstructed new audio signals based on these estimations. We dynamically scaled the amplitude envelope of the reference audio signal $p_{1}(t)$ at each 1/3-octave frequency band using the directivity factors $\bm{D}^\prime_\tau$ estimated at each frame $\tau$ by the MLP. This yielded a new, estimated set of audio signals $\bm{p^\prime}$ at coordinates $\bm{\Omega}_{\Gamma}$. Figure \ref{fig:ltas}b shows the LTAS of the estimated audio signals in front of, to the left of, and behind the speaker. Next, we compared the estimated LTAS-DI of each signal $p^\prime_{g}(t)$ to the analytical LTAS-DI calculated in Section \ref{sec:measure} (shown in Figure \ref{fig:ltas}b). This comparison enabled us to gain a perceptual understanding of the accuracy of the reconstructions. This comparison is shown in Figure \ref{fig:reconstruction-polar}. The polar plots each show a 2-D equatorial slice at $\phi=\pi/2$ of the estimated LTAS-DI predicted by the MLP and LSTM models as compared to the analytical LTAS-DI. Both models produced a fairly accurate reconstruction of the directivity pattern along the azimuthal plane for frequencies up to 1 k Hz. This observation, particularly regarding the LSTM model, suggests that the majority of directivity patterns did not exhibit significant deviations from the mean within this frequency range. However, as the frequency increased beyond 1 k Hz, the estimation of the rear lobes became less precise. The LSTM struggled to approximate the rear lobes, presumably due to its tendency to predict the mean pattern across all frequency bins. Similarly, the MLP failed to capture the details of the rear lobes but performed slightly better in estimating the side lobes.

\subsection{Influence of Objective Function Modifications}
The MLP was trained to predict target directivity patterns and objective function variations as outlined in Section \ref{sec:architectures}. Table \ref{tab:results-loss} summarizes the results of these experiments. It is not immediately clear that the domain-inspired objective function modifications improve model performance. Specifically, the sign-weighted loss modification $d_{sw}$ did not appear to enhance model performance. The two target modifications, $\bm{y}_{A}$ and $\bm{y}_{D}$, produced similar levels of accuracy. 
In the following section, we discuss the benefits of decoding to the directivity factor from the spherical harmonics domain before or after training.

\begin{table}[tb!]
\centering
\caption{Directivity Factor Reconstruction Error $\Delta \bm{D}$ on the validation set after applying objective function modifications from Table \ref{tab:objective-funcs}.}
\label{tab:results-loss}
\begin{tabular}{@{}lcccc@{}}
\toprule
 & & \multicolumn{1}{r}{Target} & & \\ \midrule
\multicolumn{1}{l|}{Obj. Func.} & $y_D$ & $y_{D,e}$ & $y_A$ & $y_{A,e}$ \\ \midrule
\multicolumn{1}{l|}{$d(\cdot)$} & 0.198 & 0.198 & 0.200 & \textbf{0.196} \\
\multicolumn{1}{l|}{$d(y_{fw}, y^\prime_{fw})$} & 0.199 & 0.197 & 0.203 & 0.199 \\
\multicolumn{1}{l|}{$d_{sw}(\cdot)$} & 0.194 & \textbf{0.192} & \textbf{0.197} & 0.199 \\
\multicolumn{1}{l|}{$d_{sw}(y_{fw}, y^\prime_{fw})$} & \textbf{0.192} & 0.197 & 0.202 & 0.196 \\ \bottomrule
\end{tabular}
\end{table}

\section{Summary and Future Work}\label{sec:Discussion}



Our framework provided us with a flexible way to capture and reconstruct high-resolution speech directivity patterns produced by unconstrained speakers using spherical harmonic encodings and simple neural networks. We demonstrated that this method can produce a temporally-varying directivity pattern from an egocentric viewpoint when provided a sound source contained within an enclosed capture system. The flexibility of the proposed encoding process allows for future directivity measurement experiments to occur within a natural communication ecology. For example, directivity patterns at multiple radii can be calculated easily and quickly using the steps described in Sections \ref{sec:real2virtual} and \ref{sec:df_theory}. Additionally, a focused study on the effects of body movement and physicality on directivity can be performed.

We also explored using simple, lightweight models such as the MLP and LSTM for estimating a dynamic high-resolution directivity pattern given a few input speech signals near the mouth. Our initial estimation experiments showed promising results for real-world applications. We investigated different domains for representing the directivity pattern. Our findings suggest that using an intermediate spherical harmonic representation yields similar accuracy to learning the directivity factor directly. Both learning paradigms have advantages: spherical harmonic encoding offers flexible decoding and easy incorporation into a larger system where this encoding is needed, while direct learning provides physical interpretability within the training process. The objective function could be optimized for perceptual accuracy during training using the distance metric as a proxy for the reconstruction error between the estimated and ``target" analytical directivity patterns. We observed a very small improvement when the model aimed to learn deviations from the expected value of the directivity pattern. With the acquisition of more spatial speech recordings, this modification may be more successful in capturing individual differences in a speaker's directivity pattern.

Due to COVID-19 protocols in place during our research, we were limited in the amount of data we could collect. We used two minutes of continuous speech from a male speaker that was recorded before access to the capture chamber was restricted. Therefore, we limited the scope of our experiments to a single speaker and used simple machine learning architectures suitable to a small dataset. However, the next critical step in this work will be to expand the scope of the experiments beyond a single speaker by collecting additional unconstrained spatial speech recordings, as described in Section \ref{sec:experiments}. Increasing the dataset size will enable further experimentation with model architectures. In this study, we did not find that the LSTM outperformed the traditional linear regression models or simple MLP with our small dataset. However, we hypothesize that this may change with an increase in dataset size, as an expressive architecture that leverages either recurrence or attention (e.g., transformers \cite{Vaswani2017}) is often more successful in modeling variance over time. Increasing the dataset size will also increase the distribution of the speech signals' spectral composition per frame, yielding a more complete representation space for the model to learn and a balanced phonetic distribution among all training, validation, and inference partitions. Additionally, collecting data from more speakers will allow for further exploration of the variation in directivity between speakers.

We intend to replace the input representation (depicted in Figure \ref{fig:grid-designs}c) with real speech signals captured by microphones on an HMD in order to validate the regressive mapping between the HMD microphone signals and high-resolution spherical reconstructions. Furthermore, we are currently conducting a tangential study to obtain additional directivity patterns using a physical spherical microphone array. These directivity patterns will provide measurements for the high and low elevation regions that could not be analyzed with the current microphone positioning in the capture chamber (refer to red markers in Figures \ref{fig:capture chamber} and \ref{fig:grid-designs}).

Finally, integrating speech directivity into a virtual sound propagation ecosystem in future research will offer valuable feedback for improving the framework proposed in this article. For instance, it remains unclear whether the larger estimation errors occurring at the rear and side lobe nulls have a perceptually salient impact on vocal presence. Gaining an understanding of how details of a directivity pattern propagate through a virtual environment would provide valuable insights into perceptual error thresholds and corresponding accuracy requirements. These insights can be used to make improvements to machine-learning designs. They can also enhance the utility and effectiveness of domain-inspired cost functions such as the ones proposed in this article.

\section{Conclusion}\label{sec:conclusion}

In this article, we presented a two-stage system for capturing and reconstructing high-resolution speech directivity patterns within a natural communication environment. Our results demonstrate that we can measure speech directivity patterns dynamically at high resolution across the frequency spectrum of a speech signal. Our preliminary results for a single subject showed that we can use limited speech information captured near the speaker's mouth to build a lightweight regressive model that reconstructs dynamic directivity patterns at the same detailed resolution. We highlighted the advantages of using a spherical harmonic representation in the encoding and modeling processes. Furthermore, we recommended design choices for future modeling experiments, proposed improvements to our framework, and emphasized the importance of speech directivity within the larger sound propagation pipeline required for a natural virtual communication experience.

\section*{Acknowledgements}
The authors thank the Audio Team at Meta Reality Labs Research for feedback and discussion during the research and implementation process.

\section*{Ethics approval and consent to participate}
All participants provided written informed consent for the use of their data for this study.

\section*{Availability of data and materials}
The datasets generated and analysed during the current study are not publicly available due confidentiality agreements between research collaborators and study participants. Access to the data may be granted upon reasonable request subject to negotiated terms between Meta Reality Labs and the requesting party.

\bibliographystyle{IEEEtran} 
\balance
\bibliography{biblio}

\end{document}